# Comparative Analysis of Multi-Omics Integration Using Advanced Graph Neural Networks for Cancer Classification

Fadi Alharbi [1], Aleksandar Vakanski [1,*], Boyu Zhang [1], Murtada K. Elbashir [2] and Mohanad Mohammed [3]

Affiliation: [1]*College of Engineering, Department of Computer Science, University of Idaho, Moscow, ID 83844, USA;* alha5622@vandals.uidaho.edu, vakanski@uidaho.com, boyuz@uidaho.edu

[2]*College of Computer and Information Sciences, Department of Information Systems, Jouf University, Sakaka, Aljouf 72441, Saudi Arabia;* mkelfaki@ju.edu.sa

[3]*School of Mathematics, Statistics and Computer Science, University of KwaZulu-Natal, Pietermaritzburg, Scottsville, 3209, South Africa;* mohanadadam32@gmail.com

* Correspondence: vakanski@uidaho.com

## Abstract

Multi-omics data is increasingly being utilized to advance computational methods for cancer classification. However, multi-omics data integration poses significant challenges due to the high dimensionality, data complexity, and distinct characteristics of various omics types. This study addresses these challenges and evaluates three graph neural network architectures for multi-omics (MO) integration based on graph-convolutional networks (GCN), graph-attention networks (GAT), and graph-transformer networks (GTN) for classifying 31 cancer types and normal tissues. To address the high-dimensionality of multi-omics data, we employed LASSO (Least Absolute Shrinkage and Selection Operator) regression for feature selection, leading to the creation of LASSO-MOGCN, LASSO-MOGAT, and LASSO-MOTGN models. Graph structures for the networks were constructed using gene correlation matrices and protein-protein interaction networks for multi-omics integration of messenger-RNA, micro-RNA, and DNA methylation data. Such data integration enables the networks to dynamically focus on important relationships between biological entities, improving both model performance and interpretability. Among the models, LASSO-MOGAT with a correlation-based graph structure achieved state-of-the-art accuracy (95.9%) and outperformed the LASSO-MOGCN and LASSO-MOTGN models in terms of precision, recall, and F1-score. Our findings demonstrate that integrating multi-omics data in graph-based architectures enhances cancer classification performance by uncovering distinct molecular patterns that contribute to a better understanding of cancer biology and potential biomarkers for disease progression.

## 1. Introduction

Gene expression is a fundamental cellular process that governs the translation of genetic information encoded in DNA into functional proteins, which in turn defines a cell's functionality and phenotype [1]. Recent advancements in gene expression analysis have significantly enhanced our understanding of the molecular pathology of various diseases, including cancer [2]. Particularly, the advent of high-throughput sequencing technologies has enabled genome-wide analysis of gene expression patterns, providing critical insights into tumorigenesis and cancer progression [3]. Moreover, the integration of gene expression data with other molecular datasets has proven effective for biomarker characterization, facilitating early cancer diagnostics and treatment methods selection [4]. By incorporating multiple layers of omics data, methods based on multi-omics data integration enable the identification of distinct molecular subtypes that might remain undetected with single-omics analyses.

The analysis and interpretation of gene expression data have been facilitated by Machine Learning (ML)-based methods, due to the ability for pattern recognition in complex high-dimensional data. Consequently, ML models have been successfully applied to predict disease progression, differentiate cancer subtypes, and identify therapeutic targets with high levels of accuracy [5]. For instance, ML methods based on Random Forest, Support Vector Machines, and Neural Networks have been employed as feature selection tools to identify the most relevant genes within gene expression data for specific biological processes [6–8]. In addition, Deep Learning (DL) models have been essential in integrating multi-omics data by uncovering patterns and relationships within large, comprehensive datasets, thereby enabling the classification of clinically relevant subtypes to enhance prognosis and treatment effects [9]. DL-based approaches, in particular, have been extensively used to integrate gene expression data with other types of omics data, such as proteomics and metabolomics, to improve cancer prognosis through reductionist modeling [10].

Graph Neural Networks (GNNs) have recently emerged as powerful tools for analyzing data with relational structures, such as biological networks, social networks, and knowledge graphs. GNNs are designed to operate on data represented in the form of a graph, where nodes and edges model the relationships within the graph elements [11]. In the context of biological networks, GNNs have proven effective in capturing complex interactions between biological entities, e.g., in the form of protein-protein interaction networks and gene regulatory networks [12]. By representing these interactions as graphs, GNNs model the dependencies and hierarchical structures within biological systems, providing important insights into cellular processes. Recent emerging graph-based network architecture, such as Graph Convolutional Networks (GCNs), Graph Attention Networks (GATs), and Graph Transformer networks (GTNs), have further enhanced the performance of GNNs, enabling more accurate and interpretable outcomes [13]. GCNs extend the concept of convolution from traditional grid-based data, such as images, to graph structures, enabling the capture of information from a node's neighbors and creating a localized graph representation around a node [14]. This makes GCNs especially effective for tasks such as node classification and link prediction, where relationships between neighboring nodes are particularly important. GATs build on this concept by incorporating an attention mechanism, allowing the model to assign different weights to neighboring nodes for learning from complex graphs [15]. The attention mechanism enables GATs to concentrate on the important parts of the graph, improving performance on tasks involving heterogeneous graphs or graphs where certain connections are more significant. Graph Transformer Networks (GTNs) introduce transformer network architectures into graph learning, providing the ability to handle long-range geometric dependencies within the graph [16]. GTNs are particularly helpful for graph-level prediction tasks, as they focus on learning global features across the entire graph. Together, GCNs, GATs, and GTNs graph-based architectures contribute to advanced modeling of complex data structures enabling more accurate solutions across a range of tasks.

This study provides an empirical evaluation of the performance of GCNs, GATs, and GTNs for classifying 31 types of cancers and normal tissues. In addition, the focus of the study is on evaluating the capacity for integrating multi-omics data, including different combinations of mRNA (messenger-RNA), miRNA (micro-RNA), and DNA methylation data. Our approach also applies LASSO (Least Absolute Shrinkage and Selection) regression for feature selection. Therefore, we dubbed the compared approaches LASSO-MOGCN (Multi-Omics Graph Convolutional Network), LASSO-MOGAT (Multi-Omics Graph Attention Network), and LASSO-MOGTN (Multi-Omics Graph Transformer Network). For modeling the relationships between the variables in these graph-based architectures, we used both gene correlation matrices and protein-protein interaction (PPI) networks to capture biological interaction and interdependence. The comparison reveals a notable difference in the performance between single-omics data and multi-omics data, where the use of single omics data generally resulted in lower performance compared to the integration of more than one omics data type. For example, LASSO-MOGAT achieved an accuracy of 94.88% when it uses DNA methylation alone as input, 95.67% accuracy for integrating mRNA and DNA methylation, and 95.90% accuracy based on the integration of mRNA, miRNA, and DNA methylation data. The improvement in the model performance highlights the importance of multi-omics integration in capturing biological signals by leveraging complementary information from different data types. Among the three approaches, LASSO-MOGAT architecture provides the best performance for both single-omics and multi-omics data.

Unlike previous related works, this study compares three graph-based models using multi-omics data (mRNA, miRNA, DNA methylation) to classify 31 different types of cancer and normal tissues. While graph-based methods have recently been widely applied for cancer classification [17–21], few studies have focused on systematically comparing modern graph-based architectures and their performances across diverse omics data. For instance, Ramirez et al. [22] employed graph-convolution networks (GCNs) for cancer classification using co-expressed genes and PPI networks. Nevertheless, their study did not investigate the performance of other graph architectures, such as GATs or GTNs. Schulte-Sasse et al. [23] introduced EMOGI, primarily targeting cancer gene prediction, but their work did not compare different GNN architectures for multi-omics integration. Different from prior works, our study presents a systematic comparison of GCNs, GATs, and GTNs for omics integration based on different graph structures. Furthermore, unlike existing studies, we focus on feature selection with LASSO regression while using graph-based techniques for handling complex associations between multiple omics datasets. In a previous work by our team, we introduced a multi-omics GAT framework [24], which utilizes the PPI network as a graph structure. While our previous work focused on developing a GAT model for gene expression analysis, in the work presented in this paper, we evaluate three graph models, GCNs, GATs, and GTNs, based on both correlation matrices and PPI networks as graph structures.

The key contributions of this study are as follows:

- An empirical investigation of the performance of Graph Convolutional Networks (GCNs), Graph Attention Networks (GATs), and Graph Transformer networks (GTNs) based on integrating mRNA or RNA-Seq, miRNA, and DNA methylation data for classification of 31 types of cancer and normal tissues.

- A comparative analysis between the performance of two graph structures based on correlation matrices and protein-protein interaction (PPI) networks.
- The performance of LASSO-MOGAT utilizing correlation matrices demonstrated state-of-the-art performance in comparison to existing neural network models using multi-omics data integration for cancer classification.

## 2. Related Work

In recent years, studies integrating multiple omics data sets have exhibited potential in moving forward cancer research by providing an improved understanding of the disease process. Computational models based on deep learning and attention-based architectures were applied to subtype cancer and enhanced model interpretability and identify biomarkers. However, challenges such as the models' applicability for other omics data, model interpretability, and clinical applicability, as well as challenges with regard to analyzing multiple omics data with high dimensions persist.

Recently, Xiao et al. [25] introduced Multi-Prior Knowledge Graph Neural Network (MPKGNN) for cancer molecular subtype classification. The authors demonstrated that the incorporation of multiple Prior Knowledge Graphs in GNNs increases the volume of omics data that it is capable of classifying with high accuracy. Chatzianastasis et al. [26] developed Explainable Multilayer Graph Neural Network (EMGNN) for cancer gene prediction using gene interaction graphs and multi-omics data. This work identifies more than one interaction network for improved performance, however the ability for generalization to high dimensions and large-scale datasets remains unexplored.

GCNs have also been used for cancer data analysis by integrating patient similarity networks with other omics data. For instance, Moon et al. [27] developed the Multi-Omics Module Analysis (MOMA) model, and Li et al. [28] introduced the Multi-omics Graph Convolutional Network (MoGCN). However, these works do not account for the various aspects of multiple omics data integration, especially the inter-omics associations essential in the classification of subtypes of cancer. Specifically, in MoGCN by Li et al. [28] GCNs are combined with autoencoders for feature learning, but Similarity Network Fusion formulates a significant portion of interactions across multiple omics layers which are important and can be neglected.

Wang et al. [29] proposed Multi-Omics Graph Convolutional Networks (MOGONET), which includes a cross-omics correlation learning process. Although this study is more exhaustive, it also has drawbacks regarding the model interpretability, as it is unclear how individual biomarkers are chosen for a particular disease. Likewise, Schulte-Sasse et al. [23] and Peng et al. [30] target enhancing biomarker discovery, however, the inability to clearly explain these models hampers their utility in precision medicine. This challenge has been mitigated by the use of attention mechanisms. Zhang et al. [31] and Guo et al. [32] applied attention-based GCNs for cancer classification with improved model explanations of cancer classification and patient biomarkers. Nevertheless, these models pose challenges in terms of scalability, generalizability, and robustness for complex multi-omics datasets. The use of attention mechanisms may also lead to overfitting in a scenario where the number of samples used is significantly smaller compared to the number of features.

Graph Transformer Networks (GTNs) are also promising architectures as they are capable of capturing long-range dependencies in complex biological data. Kaczmarek et al. [33] and Wang et al. [34] applied GTNs for cancer sample classification, combining multi-omics data with miRNA and mRNA sequencing data. Although these models offer better interpretability, especially in the identification of key biological pathways, they perform poorly in terms of accuracy compared to the traditional deep learning methods, thus a better combination of omis fusion techniques is required. There has also been extensive application of autoencoders for dimensionality reduction in multi-omics data. Although the works by Zhang et al. [13] and Chai et al. [35] achieved high cancer classification accuracy, autoencoders are generally considered a black-box solution that lacks explanation, which can critically impact the utility of these algorithms. Additionally, due to the lack of ability to decode the intricate relations of different omics, autoencoders have shortcomings in comparison to feature learning with graph-based methods.

The study by Khemani et al. [36] explores certain GNN models, such as graph attention networks (GATs), graph convolution networks (GCNs), and GraphSAGE that are currently widely employed in many different applications. Similarly, Lachi et al. [37] investigated scalable multi-graph pretraining. These studies addressed some of the prevalent challenges for generalizability and performance gain in graph learning. Although these works offer a comparison of advanced computational models, they fail to answer questions about the optimal level of balance between performance and interpretability in multi-omics data analysis. Besides, research works aiming at solving problems on a larger scale, such as Zhao et al. [38] and Yang et al. [39], struggle with algorithmic issues of GNN models to process large and complex multi-omics data sets. These models also fail to learn long-range dependencies in the data and between different omics features that are critical to comprehending tumor progression and therapeutic response.

In conclusion, while efforts using graph-based architecture for cancer classification and biomarker identification have yielded promising results, several limitations remain. Prior approaches tend to prioritize performance or interpretability, nonetheless, models that effectively balance both characteristics are relatively scarce, particularly in the context of large, heterogeneous multi-omics data integration. Furthermore, the integration of omics data types including RNA-Seq, miRNA, and DNA methylation is still challenging because most approaches do not capture the intricacies of the relationship between the data types. Consequently, there is a growing demand to develop novel approaches that are capable of unifying multi-omics data in interpretable and computationally efficient frameworks, and with a high prediction power.

## 3. Materials and Methods

### 3.1 Data Collection

The multi-omics data for the different cancers used in this study were downloaded from the Pan-Cancer Atlas [40] using the GDC query tool from the TCGAbiolinks library [41]. The Genomic Data Commons is a project set up by the National Cancer Institute which is dedicated to providing a central database that can be used by researchers wishing to carry out cancer genomic studies.

The GDCquery function requires several parameters: project, legacy, data.category, data.type, and sample.type. The project parameter determines the TCGA project from which data should be retrieved. The TCGA consists of multiple cancer research initiatives, each focusing on a specific subtype or aspect of cancer. For this study, the project parameter was set to "TCGA-*", encompassing all 33 TCGA projects (cancer types in addition to the normal tissues). The legacy parameter was set to True, indicating that the query should retrieve original, unaltered data from the TCGA Data Portal's legacy repository. The data.category parameter identifies the relevant category for the project, which in this case included transcriptome profiling for mRNA or RNA-Seq, miRNA datasets, and the DNA Methylation category for methylation data. The data.type argument was used to filter files, specifying Gene Expression Quantification for mRNA or RNA-Seq data, miRNA Expression Quantification for miRNA data, and Methylation Beta Value for DNA Methylation data. The sample.type parameter defined the sample types to be downloaded; in our case, this was set to "Primary Solid Tumor" and "Solid Tissue Normal" to extract omics data from both tumor and non-tumor samples.

The data was arranged so that the attributes of each data type were listed along the vertical axis, or ‘rows’, whilst the samples or cases were identified along the horizontal axis, or ‘columns’. The tumor types and the number of samples included in the TCGA multi-omics data (mRNA, miRNA, and DNA methylation) included in this study are depicted in Figure 1.

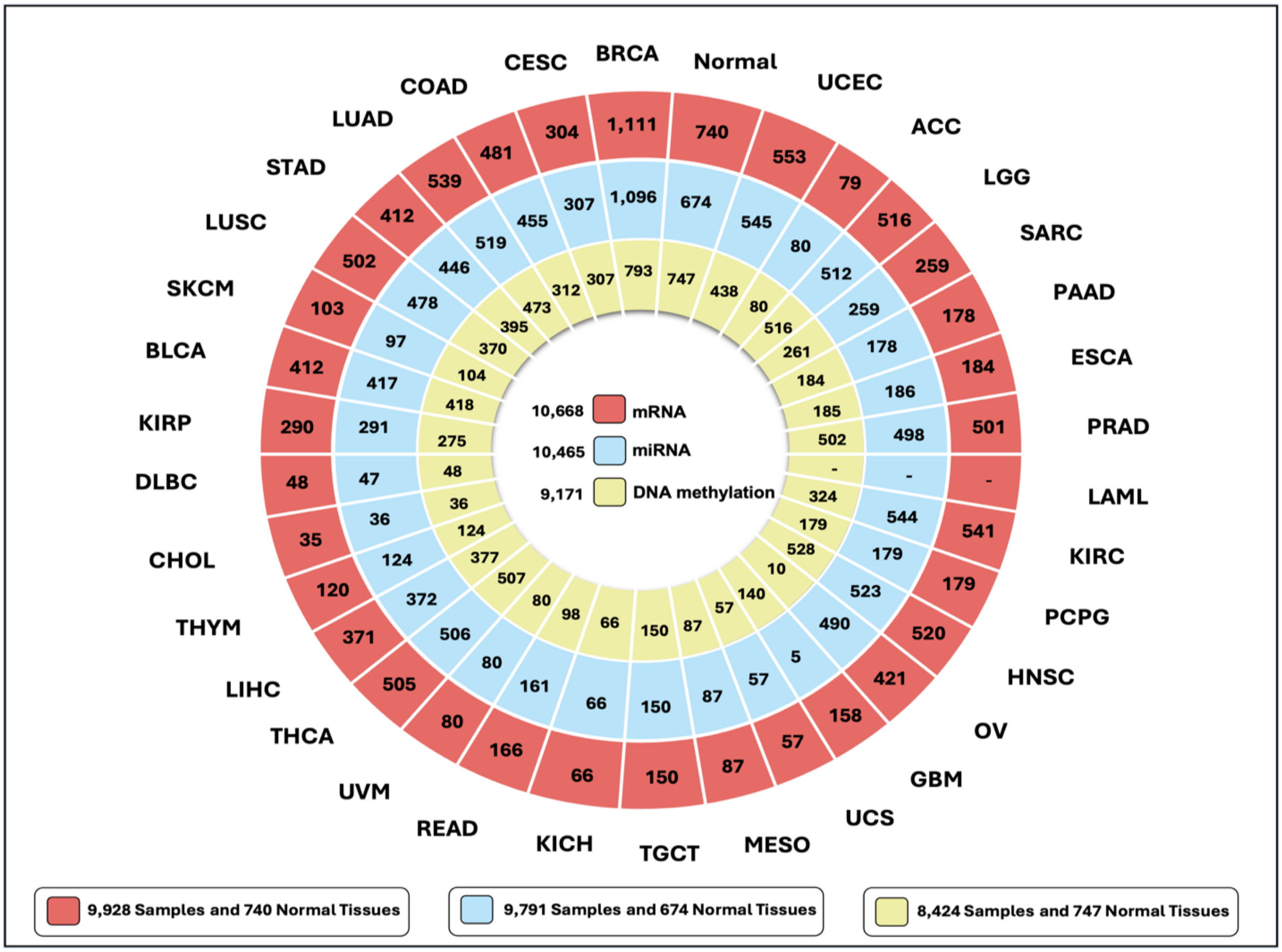

**Figure 1.** Tumor types including the number of samples and normal tissues of TCGA Multi-omics data (mRNA, miRNA, and DNA methylation) used in the analysis.

## 3.2 Data Preprocessing

The aim of this study is classification of 31 types of cancer employing RNA-Seq, miRNA, and DNA methylation [41]. We normalized the RNA-Seq data and conducted differential gene expression (DGE) analysis. The significant genes were selected based on the adjusted *p*-value for which we set the threshold to 0.05 [42]. For miRNA and DNA methylation data, the limma R package was used to identify significant features with the cut-off of adjusted *p*-value set to 0.05 [43]. LASSO regression was used afterward to further reduce the number of features where those with non-zero coefficients were retained [44].

### 3.2.1 Differential Gene Expression (DGE) Analysis

Differential gene expression (DGE) profiling is a method widely used in genomics to compare the levels of gene expression in a given organism under different conditions or environments, for example, treatment versus control, or normal versus cancer, etc. [45]. This helps in the understanding how genes are regulated and their actions depending on the environmental conditions, and a host of other functions. In our analysis for the current study, we performed differential gene expression on the mRNA data using the DESeq2 R package. This method takes a general linear model of the count data for every gene, using negative binomial distribution to capture both biological variation and overdispersion. The values of the estimated log fold changes were tested for their significance by the Wald test and the genes were called differentially expressed based on the *p*-values obtained from the Wald statistic. For the purpose of pinpointing genes that are likely to be implicated in the biological processes being studied, we set the *p*-value threshold to 0.001.

### 3.2.2 LIMMA Model

We applied the LIMMA (Linear Models for Microarray) package to perform the differential methylation analysis for which we fit a linear model for the methylation levels of CpG sites as a function of the experimental sample groups [46]. The dataset has 9,171 samples and 485,577 features obtained from the Human Methylation 450K (HM450) array [47]. Using the LIMMA package, we selected CpG sites significantly methylated in tumors compared to normal samples. LIMMA provides a moderated *t*-statistic for each CpG site and provides an estimate of the effect size that reflects the relative difference in methylation between the groups. The *p*-value that corresponds to the *t*-statistic shows the statistical significance of these differences. In this paper, we used a cutoff *p*-value less than 0.05 to filter the data, and thus the number of features was reduced to 13,9321 that concentrate on the most significant methylation change.

### 3.2.3 LASSO Regression Model

Lasso regression is a linear regression method with an additional term in the objective function by which the coefficients of the less important features are reduced to zero. The objective function for Lasso regression is defined as:

$$\min_{\beta}\left\{\frac{1}{2n}\sum_{i=1}^{n}\left(y_i - X_i\beta\right)^2 + \lambda\sum_{j=1}^{p}\left|\beta_j\right|\right\} \quad (1)$$

where the residual sum of squares is $\frac{1}{2n}\sum_{i=1}^{n}\left(y_i - X_i\beta\right)^2$, and the $L_1$ penalty term that encourages the model to learn sparse coefficients is $\lambda\sum_{j=1}^{p}\left|\beta_j\right|$. When adjusted by the parameter $\lambda$, Lasso regression establishes a reasonable and balanced relationship between the fit and the complexity of the model, using the estimates of the coefficients for the most significant features for the prediction of the response variable. This approach allows to obtain the final set of features to be used for further analysis that exclude less important features.

For mRNA and DNA methylation data we used the LASSO regression for feature selection and feature regularization. After applying LASSO regression, the number of mRNA features decreased from 26,768 to only 520, and the number of DNA methylation features from 139,321 to 393. The pipeline for data processing is summarized in Table 1.

**Table 1.** Pipeline for data processing. Row 1 (Original Features): number of original features, Row 2 (Differentially Expressed Genes): number of features after applying Differentially Expressed Analysis, Row 3 (LIMMA model): number of features after applying LIMMA model, Row 4 (LASSO Regression Model): number of features after applying LASSO Regression Model, Row 5 (All Tumor Samples and Normal): number of tumor samples and normal samples, Row 6 (Unique Tumor Samples and Normal): number of tumor and normal samples after removing duplicates, Row 7 (Common Samples and Features): number of tumor and normal samples and features common across all datatypes, and Row 8 (Network Nodes and Edges): number of nodes and edges for the PPI network for each datatype.

| Datatype | mRNA | miRNA | DNA Methylation |
|---|---|---|---|
| Original Features | 60,660 | 1881 | 485,577 |
| Differentially Expressed Analysis | 26,768 | - | - |
| LIMMA Model (Selected Features) | - | - | 139,321 |
| LASSO Regression Model (Selected Features) | 520 | - | 393 |
| All Tumor Samples and Normal Tissues | 10,668 | 10,465 | 9,171 |
| Unique Tumor Samples and Normal Tissues | 10,667 | 10,465 | 8,674 |
| Common Samples and Features | 8,464 Samples and 2,794 Features | | |
| Network Nodes and Edges | 504 Nodes and 343 Edges | | |

### 3.3 Multi-Omics Data Integration

We merged all omics data pertaining to each sample into a single record, integrating mRNA or RNA-Seq, miRNA, and DNA methylation data using the sample ID as a common value. The inner join merging operation was used on the sample ID of the three datasets and only samples with the complete data in all the three omics data were retained. This approach helped to address the problems of missing data, as the final integrated dataset contains only samples with complete data in all three omics datasets. For some cancer types, there were missing data for one or more omics data; for instance, 'TCGA_LAML' cancer type had no RNA-Seq data, and 'TCGA_GBM' had no miRNA data. For further analysis, we omitted cancer types for which omics data was incompletely represented. The final dataset comprises 8,464 samples involving 2,794 omics features and covers 31 types besides the normal tissues. This dataset is therefore suitable for additional analyses with a view of establishing diverse molecular mechanisms that can be associated with various forms of cancer. The preprocessing steps and data integration are illustrated in Figure 2.

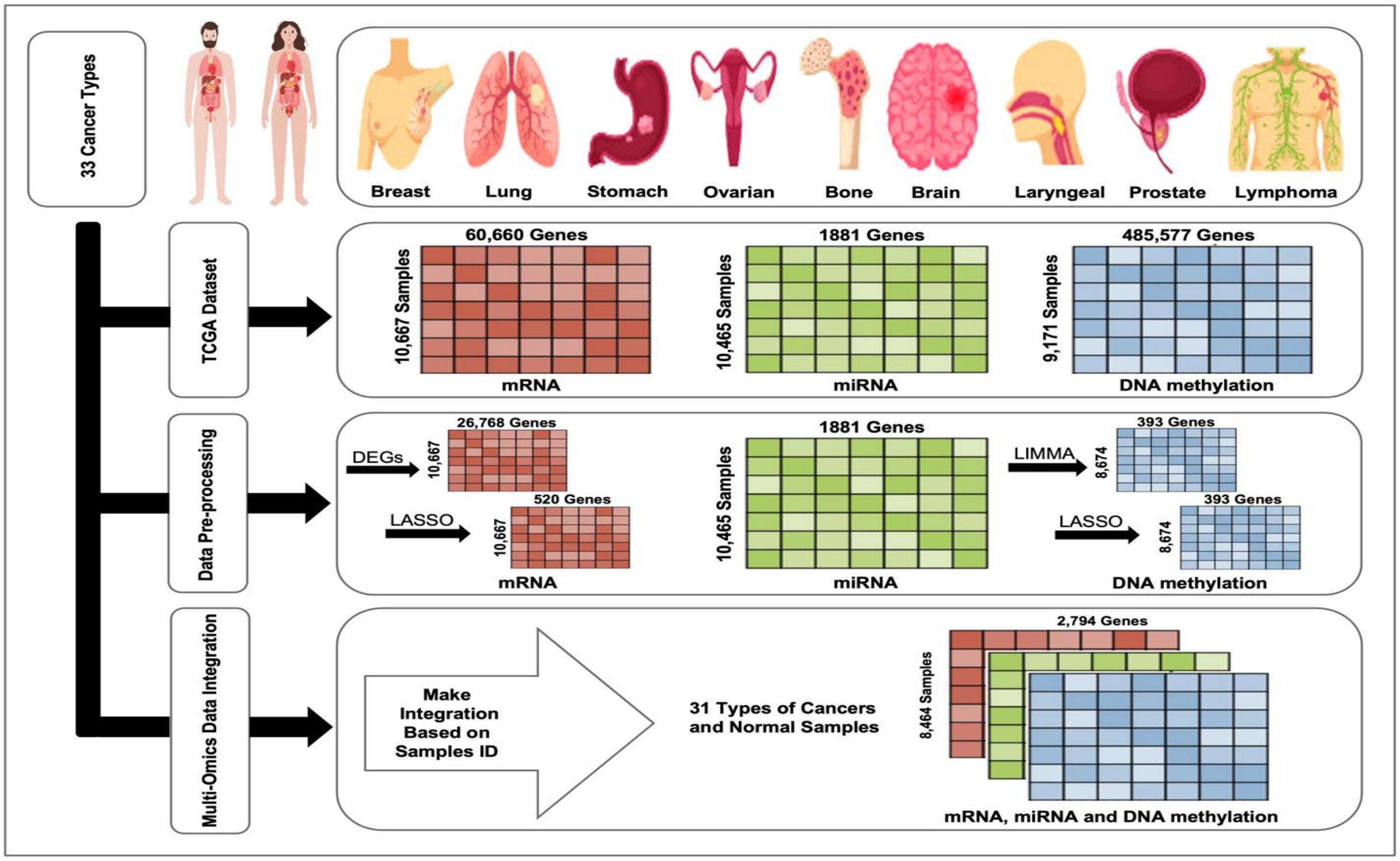

**Figure 2.** Preprocessing steps and data integration: First, omics data (mRNA, miRNA, and DNA methylation) were obtained from the Pan-cancer Atlas using the TCGAbiolinks library. Next, differential expression analysis (DEG) and LASSO regression were applied to mRNA data, while LIMMA and LASSO regression were applied to DNA methylation data. Subsequently, mRNA or RNA-Seq, miRNA, and DNA methylation data were integrated based on the sample ID using an inner join operation.

## 3.4 Graph Structures

### 3.4.1 Correlation Matrices

For the graph construction, we employed two types of graph structures: correlation matrices and the molecular topology of PPI networks. The correlation matrix for each omics data type was computed to define the graph structure using the correlation coefficient to measure the relationship between pairs of features. This correlation matrix to define the graph structure was obtained by taking a cutoff at 0.7. The correlation coefficient between pairs of features is as follows:

$$corr(X_i, X_j) = \frac{cov(X_i, X_j)}{\sigma_{X_i}\sigma_{X_j}} \tag{2}$$

### 3.4.2 Protein-Protein Interaction (PPI) Network

The protein-protein interaction (PPI) network depicts the relationship existing between proteins in a cell [48,49]. Proteins interact to conduct numerous biological processes including metabolism, gene regulation, and signaling, and therefore PPI networks are crucial for describing pathophysiological processes like cancer, and other commonplace cellular processes [50, 51]. The PPI network was constructed based on the corresponding genes from the STRING database. The String database is a valuable resource for both predicted and experimentally validated PPI data [52]. In PPI, the nodes refer to proteins and edges refer to the known relationships between them [48]. This data involves the identification of the proteins involved as well as the strength of the interaction between them. The nodes and edges were coded as tensors, and afterward were fed into the different Graph Network architectures as a part of the structure of the graph. In this graph-based representation, each patient is described by the topology of the PPI network incorporating nodes with multi-omics features obtained from DNA methylation, miRNA, and mRNA. While the topology is identical across the patients, the node features may be different indicating the unique molecular profiles of cancer for the specific patient. This approach allows for elaborate and personalized assessment of cancer-related molecular signaling.

## 3.5 Graph Neural Network Architectures

### 3.5.1 Graph Convolutional Networks (GCNs)

The family of the GCNs generalizes the concept of convolution common for traditional grid data to graph data. In GCNs, a node aggregates information from its neighboring nodes to compute its features, and thus the model captures the localized spatial structure of a graph. This is especially helpful for node classification, graph classification, and link prediction. For instance, GCNs may study the protein-protein relations or merge various types of omics data for cancer typing or the evolution of a disease. The analysis of the features of GCNs reveals that for the tasks in which the graph structure presents a high level of relevance, the algorithms can capture interactions and dependencies, which provides high flexibility in the use of ML algorithms and data analysis. The layer-wise propagation rule for the GCNs using the correlation matrix or the PPI structure can be defined as [12]:

$$H^{(l+1)} = \sigma(D^{-\frac{1}{2}} A D^{-\frac{1}{2}} H^{(l)} W^{(l)}) \quad (3)$$

$H^{(l)}$ stands for the feature matrix at layer $l$, $W^{(l)}$ is a weight matrix, $A$ is an adjacency matrix, $D$ is the degree matrix, and $\sigma$ is an activation function [14].

### 3.5.2 Graph Attention Networks (GATs)

GATs are neural network architectures that are capable of learning data representations that are arranged in the form of graphs by applying an attention mechanism to the nodes of the graph. While GCNs operate by taking an average of all features from the neighboring nodes, GATs also include information from neighbors, where using an attention layer permits the model to decide the weighting of the parameters based on neighboring nodes. This is accomplished through weighting of the graph edges where during the updating of the features of a node, the model pays special attention to the most relevant neighbors. GATs are particularly suitable for node classification, graph classification, and link prediction problems if the connections between nodes are not all similar and all relations are not equally significant. GATs are broadly in bioinformatics, where dense networks like protein-protein interactions enhance the predictions in cases like cancer classification. The capacity to decide dynamically the contribution of each neighbor makes GATs an effective model for learning relatively expressive patterns from graph-based data. The employed attention mechanism for the feature update rule of a node is given as follows [53]:

$$h_i = \sigma(\sum_{j\in\mathcal{N}(i)} \alpha_{ij}\, Wh_j) \tag{4}$$

and the attention coefficients $\alpha_{ij}$ are computed using the formula [15]:

$$\alpha_{ij} = \frac{\exp\,(LeakyReLU(a^T[Wh_i \| Wh_j]))}{\sum_{k\in\mathcal{N}(i)} \exp\,(LeakyReLU(a^T[Wh_i \| Wh_k]))} \tag{5}$$

### 3.5.3 Graph Transformer Networks (GTNs)

GTNs are one of the state-of-the-art neural networks to process graph-structure data. Utilizing the attention mechanisms of Transformers, they can learn long-range dependencies, as well as embedded interactions inside the graph. In contrast to the original GNNs that use only local neighborhood aggregation, GTNs are able to represent more complex relations since the model learns which nodes of the graph should attend to each other regardless of their distance in the graph. GTNs are effective for node classification, graph classification, and link prediction, since the relationships between nodes are more intricate and non-local engagement is more important. In bioinformatics, GTNs have been used for the analysis of multi-omics data in which they are capable of finding out important patterns and relationships between various levels of biological information consisting of transcriptions, methylations, and proteins and their interactions. As a result, GTNs offer an alternative methodology for graph processing while retaining both local and global dependencies inherent in graph-structured data. The update rule for the transformer layer in GTNs is as follows [54]:

$$H^{(l+1)} = TransformerLayer(H^{(l)}, A) \tag{6}$$

where incorporating self-attention mechanisms [14] described by:

$$Attention(Q, K, V) = softmax\left(\frac{QK^T}{\sqrt{d_k}}\right)V \tag{7}$$

# 4. Experimental Evaluation

## 4.1 Experimental setup

In this study, the multi-omics data was standardized, and two separate graph structures were employed: one based on PPI network and another on gene correlation. The PPI network was formulated based on interaction data derived from the String database which shows proteins as nodes and their interactions as edges. Each of the graphs was converted to tensors and the multi-omics data were combined with the nodes of the graph to create data objects that can be fed into the graph architecture network independently. This allows comparing the efficiency of each graph structure in identifying the molecular connections and dependencies in the data. The model architectures of GAT, GTN, and GCN were evaluated with a 5-fold cross-validation approach.

The GAT architecture consists of four GATConv layers, which were configured with the following parameters: the first layer has 1024 output channels and 8 attention heads. The second layer cuts down the dimensions to 512 channels and has 4 attention heads. The third layer has 256 channels and two attention heads, and the final layer has 32 channels with one attention head. Batch Normalization and Leaky ReLU activation are used after each layer. We used 0.5 dropout rate to prevent overfitting. We used the Adam optimizer with an initial learning rate of 0.001. A learning rate scheduler (ReduceLROnPlateau) is used to adjust the learning rate dynamically based on the validation accuracy. Negative Log Likelihood Loss, i.e., cross-entropy, is used as a loss function. The model was trained for 100 epochs within each fold of the 5-fold cross-validation, where data was split into training and testing sets. The parameters were optimized based on the grid search approach.

The GCN model was designed with the following key parameters where the standardized features from multi-omics data are used as an input layer. The model architecture construction starts with the first GraphConv layer to which the input features are passed in and then proceeds with 2 hidden GraphConv layers. Each of the hidden layers has 128 hidden units with ReLU activation function. Also, a dropout layer with a 0.5 dropout rate is applied after each layer. The Adam optimizer with a learning rate of 0.001 is employed in the training process of 100 epochs. The parameters were optimized based on the grid search approach.

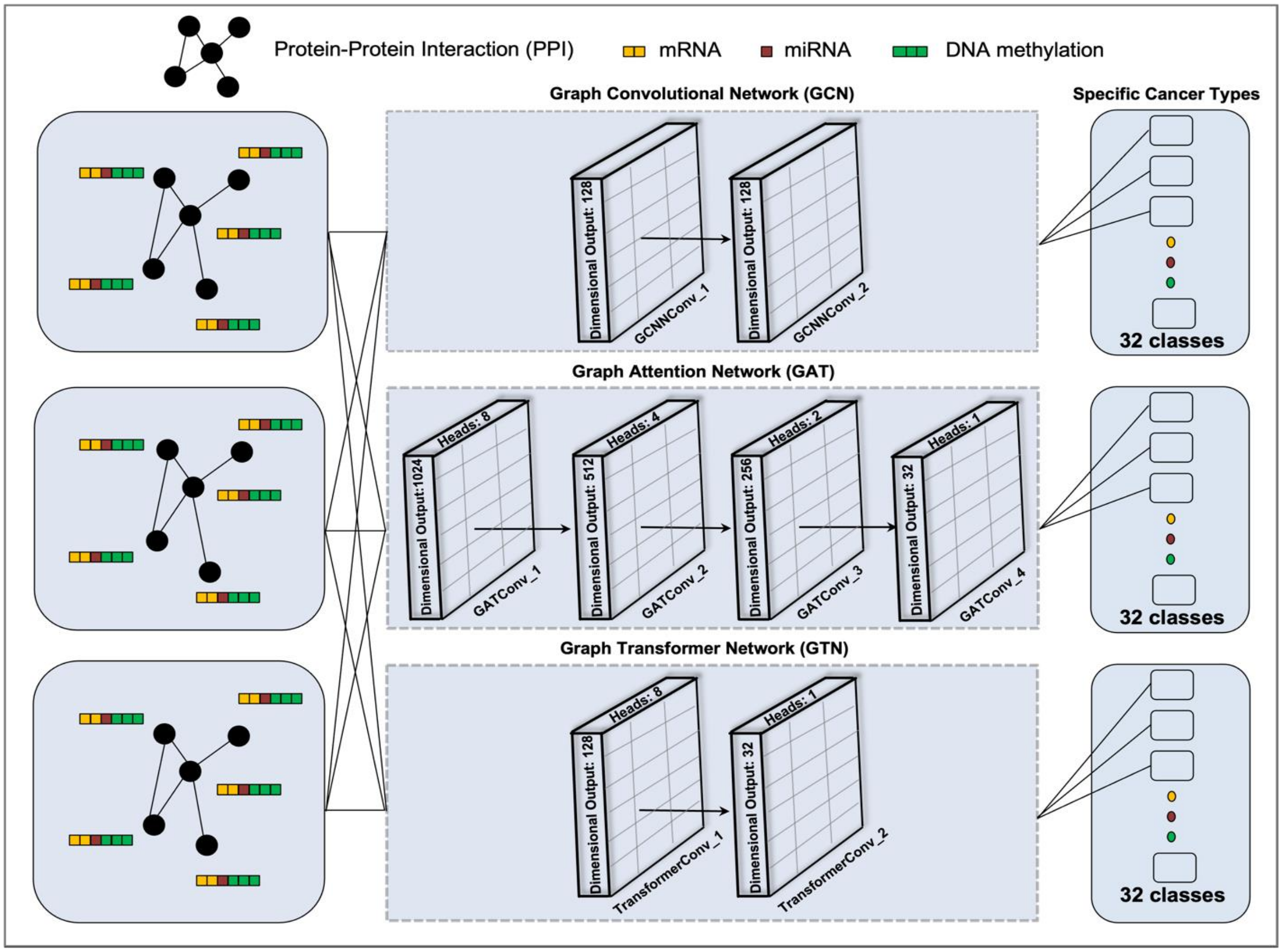

**Figure 3.** The architecture of the GCN, GAT, and GTN models for multiclass cancer classification.

The GTN architecture comprises two TransformerConv layers configured with 8 attention heads each. The first TransformerConv layer takes the multi-omics data as the input and produces 64 channels as the output. The second TransformerConv layer additionally processes the output of the first TransformerConv layer and reduces the dimensionality to match the number of classes for final classification. Each TransformerConv layer is responsible for different portions of the input features and also collects data from adjacent nodes in the graph. We used the Adam optimizer with a learning rate of 0.001 in the training process. We used the CrossEntropyLoss as a loss function. ReduceLROnPlateau was used as a learning rate scheduler to adjust the learning rate based on the validation accuracy dynamically. We trained the model for 100 epochs.

## 4.2 Performances Metrics

All the models were coded in Python using the PyTorch Geometric library. The multi-omics dataset was split into training, validation, and test sets, and the performance of the models was assessed using a five-fold cross-validation approach. The evaluation metrics include accuracy, precision, recall, and F1-score. The accuracy is:

$$Accuracy = \frac{TP+TN}{TP+TN+FP+FN} \tag{8}$$

Since it is a multi-class classification problem, we calculate the macro-averaged precision, recall, and F1 score for the folds. The equations are as follows:

$$\text{Macro} - \text{Precision} = \frac{1}{C}\sum_{c=1}^{C}\frac{TP_c}{TP_c+FP_c} \tag{9}$$

$$\text{Macro} - \text{Recall} = \frac{1}{C}\sum_{c=1}^{C}\frac{TP_c}{TP_c+FN_c} \tag{10}$$

$$\text{Macro} - \text{F1} = \frac{1}{C}\sum_{c=1}^{C}\frac{2\times Precision_c \times Recall_c}{Precision_c+Recall_c} \tag{11}$$

## 4.3 Experimental Results

The performance of the LASSO-MOGCN model for different combinations of multi-omics data when the correlation is used as a graph structure is presented in Table 3. The table shows that LASSO-MOGCN achieved the highest performance when applied to the combination of mRNA or RNA-Seq, miRNA, and DNA methylation. In this combination, it achieved an accuracy, precision, recall, and F1 score of 95.39%±0.003, 0.9462±0.010, 0.9374±0.006, and 0.9380±0.004, respectively. This indicates that the combination of the three omics data types is able to capture the most comprehensive biological signal. The low values of standard deviation imply stable performance across different test sets. Generally, the model performance on the combination of multi-omics data is better than its performance on single omics data, where the worst performance is achieved when it is applied to the miRNA data alone.

**Table 3.** Performance metrics of the proposed LASSO-MOGCN approach based on correlation matrices.

| Data Types | Multi-Omics Data Type | Cancer Types | Accuracy<br>Mean ± std | Precision<br>Mean ± std | Recall<br>Mean ± std | F1 Score<br>Mean ± std |
|---|---|---|---|---|---|---|
| Single Omics Data | mRNA or RNA-Seq | 31 Types of Cancer and Normal Tissues | 93.44% ± 0.005 | 0.9197 ± 0.004 | 0.9119 ± 0.007 | 0.9129 ± 0.003 |
| Single Omics Data | miRNA | 31 Types of Cancer and Normal Tissues | 87.92% ± 0.004 | 0.8611 ± 0.015 | 0.8500 ± 0.015 | 0.8521 ± 0.015 |
| Single Omics Data | DNA methylation | 31 Types of Cancer and Normal Tissues | 91.70% ± 0.008 | 0.8838 ± 0.027 | 0.8663 ± 0.021 | 0.8657 ± 0.023 |
| Multi-Omics Data | mRNA or RNA-Seq and miRNA | 31 Types of Cancer and Normal Tissues | 94.05% ± 0.003 | 0.9331 ± 0.008 | 0.9204 ± 0.013 | 0.9224 ± 0.011 |
| Multi-Omics Data | mRNA or RNA-Seq and DNA methylation | 31 Types of Cancer and Normal Tissues | 95.30% ± 0.004 | 0.9392 ± 0.009 | 0.9309 ± 0.005 | 0.9320 ± 0.003 |
| Multi-Omics Data | miRNA and DNA methylation | 31 Types of Cancer and Normal Tissues | 93.96% ± 0.002 | 0.9255 ± 0.019 | 0.9109 ± 0.008 | 0.9147 ± 0.012 |
| **Multi-Omics Data** | mRNA or RNA-Seq, miRNA and DNA methylation | 31 Types of Cancer and Normal Tissues | **95.39% ± 0.003** | **0.9462 ± 0.010** | **0.9374 ± 0.006** | **0.9380 ± 0.004** |

Table 4 shows the performance of LASSO-MOGCN on different combinations of omics data where the used graph structure is the PPI network. The best performance is achieved on the omics combination of mRNA or RNA-Seq, miRNA, and DNA methylation, where the accuracy, precision, recall, and F1 score are 95.10%±0.005, 0.9348±0.012, 0.9227±0.011, and 0.9250±0.011, respectively. This performance is lower than that achieved when the correlation is used as a graph structure, presented in Table 3. The results suggest that although the PPI network affords unique insight into the protein interactions, it may not capture the necessary interdependencies of the multi-omics characteristics as correlation matrices.

**Table 4.** Performance metrics of the proposed LASSO-MOGCN approach based on PPI Network.

| Data Types | Multi-Omics Data Type | Cancer Types | Accuracy Mean ± std | Precision Mean ± std | Recall Mean ± std | F1 Score Mean ± std |
|---|---|---|---|---|---|---|
| Single Omics Data | mRNA or RNA-Seq | 31 Types of Cancer and Normal Tissues | 93.36% ± 0.002 | 0.9175 ± 0.003 | 0.9049 ± 0.013 | 0.9059 ± 0.006 |
| Single Omics Data | miRNA | 31 Types of Cancer and Normal Tissues | 88.26% ± 0.003 | 0.8572 ± 0.015 | 0.8424 ± 0.007 | 0.8446 ± 0.009 |
| Single Omics Data | DNA methylation | 31 Types of Cancer and Normal Tissues | 94.23% ± 0.002 | 0.9205 ± 0.017 | 0.9116 ± 0.015 | 0.9129 ± 0.015 |
| Multi-Omics Data | mRNA or RNA-Seq and miRNA | 31 Types of Cancer and Normal Tissues | 93.98% ± 0.001 | 0.9275 ± 0.017 | 0.9197 ± 0.011 | 0.9201 ± 0.013 |
| Multi-Omics Data | mRNA or RNA-Seq and DNA methylation | 31 Types of Cancer and Normal Tissues | 95.35% ± 0.004 | 0.9445 ± 0.011 | 0.9291 ± 0.009 | 0.9328 ± 0.011 |
| Multi-Omics Data | miRNA and DNA methylation | 31 Types of Cancer and Normal Tissues | 93.72% ± 0.006 | 0.9170 ± 0.019 | 0.9024 ± 0.018 | 0.9058 ± 0.018 |
| **Multi-Omics Data** | mRNA or RNA-Seq, miRNA and DNA methylation | 31 Types of Cancer and Normal Tissues | **95.10% ± 0.005** | **0.9348 ± 0.012** | **0.9227 ± 0.011** | **0.9250 ± 0.011** |

Table 5 presents the results of applying LASSO-MOGAT to classify the 31 types of cancers in addition to the normal tissues based on a combination of multi-omics data and the correlation matrix as a graph structure. The model achieved the highest accuracy (94.88%) among the single omics data when using DNA methylation alone. However, the model achieved better performances when the DNA methylation is combined with other omics data, E.g., the combination of mRNA/RNA-Seq and DNA methylation shows an accuracy of 95.67%, as well this combination shows robust precision, recall, and F1 scores. The best performance is achieved when the three omics data are integrated, where the accuracy, precision, recall, and F1 scores of the model are 95.90%±0.001, 0.9438 ±0.008, 0.9445±0.009, and 0.9420±0.008, respectively.

**Table 5.** Performance metrics of the proposed LASSO-MOGAT approach based on correlation matrices.

| **Data Types** | **Multi-Omics Data Type** | **Cancer Types** | **Accuracy** Mean ± std | **Precision** Mean ± std | **Recall** Mean ± std | **F1 Score** Mean ± std |
|---|---|---|---|---|---|---|
| Single Omics Data | mRNA or RNA-Seq | 31 Types of Cancer and Normal Tissues | 93.88% ± 0.004 | 0.9169 ± 0.006 | 0.9283 ± 0.013 | 0.9205 ± 0.006 |
| Single Omics Data | miRNA | 31 Types of Cancer and Normal Tissues | 89.59% ± 0.007 | 0.8665 ± 0.009 | 0.8636 ± 0.011 | 0. 8613 ± 0.006 |
| Single Omics Data | DNA methylation | 31 Types of Cancer and Normal Tissues | **94.88%** ± 0.003 | 0.9226 ±0.011 | 0.9294 ± 0.013 | 0.9242 ± 0.011 |
| Multi-Omics Data | mRNA or RNA-Seq and miRNA | 31 Types of Cancer and Normal Tissues | 94.55% ± 0.004 | 0.9229 ± 0.014 | 0.9258 ± 0.008 | 0.9224 ± 0.011 |
| Multi-Omics Data | mRNA or RNA-Seq and DNA methylation | 31 Types of Cancer and Normal Tissues | **95.67%** ± 0.002 | 0.9316 ± 0.012 | 0.9375 ± 0.017 | 0.9331 ± 0.015 |
| Multi-Omics Data | miRNA and DNA methylation | 31 Types of Cancer and Normal Tissues | 94.81% ± 0.005 | 0.9202 ± 0.010 | 0.9248 ± 0.007 | 0.9189 ± 0.009 |
| **Multi-Omics Data** | mRNA or RNA-Seq, miRNA and DNA methylation | 31 Types of Cancer and Normal Tissues | **95.90% ± 0.001** | **0.9438 ± 0.008** | **0.9445 ± 0.009** | **0.9420 ± 0.008** |

Table 6 presents the performance of the LASSO-MOGAT technique for classifying 31 types of cancer and normal tissues based on a PPI network. Even though single omics data types like DNA methylation or mRNA yield high accuracy, the results are significantly enhanced when the omics data types are combined. This combination yields the best results with an accuracy of 95.74%, and precision, recall, and F1 scores all above 93%.

**Table 6.** Performance metrics of the proposed LASSO-MOGAT approach based on PPI Network.

| **Data Types** | **Multi-Omics Data Type** | **Cancer Types** | **Accuracy** Mean ± std | **Precision** Mean ± std | **Recall** Mean ± std | **F1 Score** Mean ± std |
|---|---|---|---|---|---|---|
| Single Omics Data | mRNA or RNA-Seq | 31 Types of Cancer and Normal Tissues | 93.67% ± 0.004 | 0.9175 ± 0.007 | 0.9185 ± 0.013 | 0.9149 ± 0.010 |
| Single Omics Data | miRNA | 31 Types of Cancer and Normal Tissues | 89.15% ± 0.003 | 0.8512 ± 0.007 | 0.8513 ± 0.007 | 0.8482 ± 0.007 |
| Single Omics Data | DNA methylation | 31 Types of Cancer and Normal Tissues | 94.61% ± 0.003 | 0.9185 ± 0.016 | 0.9188 ± 0.017 | 0.9170 ± 0.016 |
| Multi-Omics Data | mRNA or RNA-Seq and miRNA | 31 Types of Cancer and Normal Tissues | 94.11% ± 0.007 | 0.9160 ± 0.023 | 0.9158 ± 0.014 | 0.9128 ± 0.015 |
| Multi-Omics Data | mRNA or RNA-Seq and DNA methylation | 31 Types of Cancer and Normal Tissues | 95.53% ± 0.002 | 0.9395 ± 0.004 | 0.9428 ± 0.007 | 0.9399 ± 0.006 |
| Multi-Omics Data | miRNA and DNA methylation | 31 Types of Cancer and Normal Tissues | 94.80% ± 0.007 | 0.9134 ± 0.023 | 0.9154 ± 0.019 | 0.9126 ± 0.021 |
| **Multi-Omics Data** | mRNA or RNA-Seq, miRNA and DNA methylation | 31 Types of Cancer and Normal Tissues | **95.74% ± 0.003** | **0.9450 ± 0.009** | **0.9354 ± 0.011** | **0.9362 ± 0.011** |

Table 7 shows the performance of LASSO-MOGTN for classifying 31 cancer types in addition to the normal tissues. LASSO-MOGTN approach based on correlation matrix yields good performance when mRNA or RNA-Seq and DNA methylation data information is used. Notably, when using all three omics types the accuracy remains moderate and equals 94%. It is important to note that LASSO-MOGTN has good performance when combining mRNA or RNA-Seq with other omics data types, such as miRNA and DNA methylation. Therefore, mRNA or RNA-Seq data might be more beneficial for cancer classification using LASSO-MOGTN with the correlation matrix-based approach. LASSO-MOGTN might fail to take full advantage of the synergistic potential of multiple omics data.

**Table 7.** Performance metrics of the proposed LASSO-MOGTN approach based on correlation matrices.

| Data Types | Multi-Omics Data Type | Cancer Types | Accuracy Mean ± std | Precision Mean ± std | Recall Mean ± std | F1 Score Mean ± std |
|---|---|---|---|---|---|---|
| Single Omics Data | mRNA or RNA-Seq | 31 Types of Cancer and Normal Tissues | 92.53% ± 0.002 | 0.9051 ± 0.013 | 0.9014 ± 0.012 | 0.9007 ± 0.012 |
| Single Omics Data | miRNA | 31 Types of Cancer and Normal Tissues | 84.78% ± 0.006 | 0.8267 ± 0.021 | 0.8018 ± 0.009 | 0.8088 ± 0.015 |
| Single Omics Data | DNA methylation | 31 Types of Cancer and Normal Tissues | 93.43% ± 0.006 | 0.9178 ± 0.019 | 0.9017 ± 0.022 | 0.9035 ± 0.021 |
| Multi-Omics Data | mRNA or RNA-Seq and miRNA | 31 Types of Cancer and Normal Tissues | 92.46% ± 0.006 | 0.9116 ± 0.018 | 0.8989 ± 0.015 | 0.9027 ± 0.017 |
| Multi-Omics Data | mRNA or RNA-Seq and DNA methylation | 31 Types of Cancer and Normal Tissues | 95.01% ± 0.002 | 0.9419 ± 0.010 | 0.9319 ± 0.009 | 0.9347 ± 0.009 |
| Multi-Omics Data | miRNA and DNA methylation | 31 Types of Cancer and Normal Tissues | 92.72% ± 0.006 | 0.9152 ± 0.010 | 0.8991 ± 0.011 | 0.9044 ± 0.009 |
| **Multi-Omics Data** | mRNA or RNA-Seq, miRNA and DNA methylation | 31 Types of Cancer and Normal Tissues | **94.06% ± 0.007** | **0.9204 ± 0.026** | **0.9089 ± 0.024** | **0.9101 ± 0.023** |

The evaluation results for the LASSO-MOGTN using the PPI network for classifying 31 types of cancer and the normal tissues are displayed in Table 8. When using mRNA or RNA-Seq data alone, the model achieved accuracy and F1-Score of 93.37% and 0.9043 respectively. The highest performance is achieved when combining mRNA with DNA methylation, where the accuracy, precision, recall, and

F1 score were 95.29%±0.002, 0.9412±0.005, 0.9351±0.005, and 0.9361±0.005, respectively. The integration of the three omics types yielded an accuracy, precision, recall, and F1 score of 95.25%±0.004, 0.9386±0.020, 0.9302±0.015, and 0.9318±0.018. This integration further highlights the benefits of integrating several omics data in the GTN model based on the PPI network.

**Table 8.** Performance metrics of the proposed LASSO-MOGTN approach based on PPI Network.

| Data Types | Multi-Omics Data Type | Cancer Types | Accuracy<br>Mean ± std | Precision<br>Mean ± std | Recall<br>Mean ± std | F1 Score<br>Mean ± std |
|---|---|---|---|---|---|---|
| Single Omics Data | mRNA or RNA-Seq | 31 Types of Cancer and Normal Tissues | 93.37% ± 0.005 | 0.9153 ± 0.006 | 0.9040 ± 0.013 | 0.9043 ± 0.010 |
| Single Omics Data | miRNA | 31 Types of Cancer and Normal Tissues | 88.58% ± 0.003 | 0.8666 ± 0.008 | 0.8519 ± 0.012 | 0.8560 ± 0.009 |
| Single Omics Data | DNA methylation | 31 Types of Cancer and Normal Tissues | 94.38% ± 0.001 | 0.9253 ± 0.014 | 0.9151 ± 0.008 | 0.9182 ± 0.010 |
| Multi-Omics Data | mRNA or RNA-Seq and miRNA | 31 Types of Cancer and Normal Tissues | 93.97% ± 0.003 | 0.9338 ± 0.009 | 0.9225 ± 0.010 | 0.9231 ± 0.011 |
| Multi-Omics Data | mRNA or RNA-Seq and DNA methylation | 31 Types of Cancer and Normal Tissues | 95.29% ± 0.002 | 0.9412 ± 0.005 | 0.9351 ± 0.005 | 0.9361 ± 0.005 |
| Multi-Omics Data | miRNA and DNA methylation | 31 Types of Cancer and Normal Tissues | 94.37% ± 0.005 | 0.9323 ± 0.021 | 0.9177 ± 0.016 | 0.9210 ± 0.018 |
| **Multi-Omics Data** | mRNA or RNA-Seq, miRNA and DNA methylation | 31 Types of Cancer and Normal Tissues | **95.25% ± 0.004** | **0.9386 ± 0.020** | **0.9302 ± 0.015** | **0.9318 ± 0.018** |

## 5. Discussion

This study provides an empirical analysis of the performance of different graph-based architectures for classifying 31 different types of cancer in addition to normal tissues. The architectures LASSO-MOGCN, LASSO-MOGAT, and LASSO-MOGTN were built based on the two-graph structures, correlation matrices and PPI network. Different feature combinations from multi-omics data comprising DNA methylation, miRNA, and mRNA or RNA-Seq are employed. The LASSO-MOGCN model shows the best performance when using the features combination mRNA or RNA-Seq, miRNA, and DNA methylation based on the correlation matrix as graph structure. The accuracy, precision, recall, and F1-score obtained based on this model were 95.39%±0.003, 0.9462±0.010, 0.9374±0.006, and 0.9380±0.004 respectively. This suggests that the GCN method can perform very well compared to other graph-based architectures with multi-omics data in classifying the cancer types when it uses correlation matrices as a graph structure. On the other hand, LASSO-MOGCN based on PPI network shows a performance lower than when it uses the correlation matrix and it obtained an accuracy, precision, recall, and F1-score of 95.10%±0.005, 0.9348±0.012, 0.9227±0.011, and 0.9250±0.011 respectively. The performance suggests that LASSO-MOGCN performs well and this discrepancy most likely results from the GCN structure's utilization of the data correlations.

LASSO-MOGTN based on the correlation matrix shows a lower performance compared to LASSO-MOGCN where it obtained an accuracy, precision, recall, and F1 Score of 94.06%±0.007, 0.9204±0.026, 0.9089±0.024, and 0.9101±0.023, respectively. Unlike LASSO-MOGCN, LASSO-MOGTN obtained better performance with the PPI network, resulting in accuracy, precision, recall, and F1 Score of 95.25%±0.004, 0.9386±0.020, 0.9302±0.015, and 0.9318±0.018, respectively. Because of its structure or the unique characteristics of the data, the GTN model may not have been able to fully utilize the potential of correlation matrices, and it is more suited to leveraging the interactions captured in PPI networks.

LASSO-MOGAT has the best performance compared to the other models when using either the PPI network or the correlation matrix. Based on the PPI network, LASSO-MOGAT achieved an accuracy, precision, recall, and F1-score of 95.74%±0.003, 0.9450±0.009, 0.9354±0.011, and 0.9362±0.011, respectively. This is the best performance compared to the other models when using the PPI network. The performance for correlation matrices of LASSO-MOGAT includes accuracy, precision, recall, and F1-score of 95.90%±0.001, 0.9438±0.008, 0.9445±0.009, and 0.9420±0.008.

This study reaffirms the efficacy of the GAT approach with correlation matrices for multi-omics data and with the PPI networks for comprehending intricate cellular interactions. Selecting the framework and data structure proves to be influential to the classification process.

Figure 4 shows the ROC curves for cancer types of classification using LASSO-MOGAT. The AUC of the classes TCGA-BLCA, TCGA-BRCA, and TCGA-ESCA reach an AUC of 1. It can also be noted that most of the ROC curves tend to be closer to the top left-hand corner indicating a high true positive rate and low false positive rate. Only two classes, TCGA-CHOL and TCGA-READ, show slightly lower AUC of 0. 97 and 0. 95, respectively, meaning that our approach can classify these types with less accuracy. The high AUC score obtained in all classes indicates that the proposed model is capable of differentiating between different cancer types well.

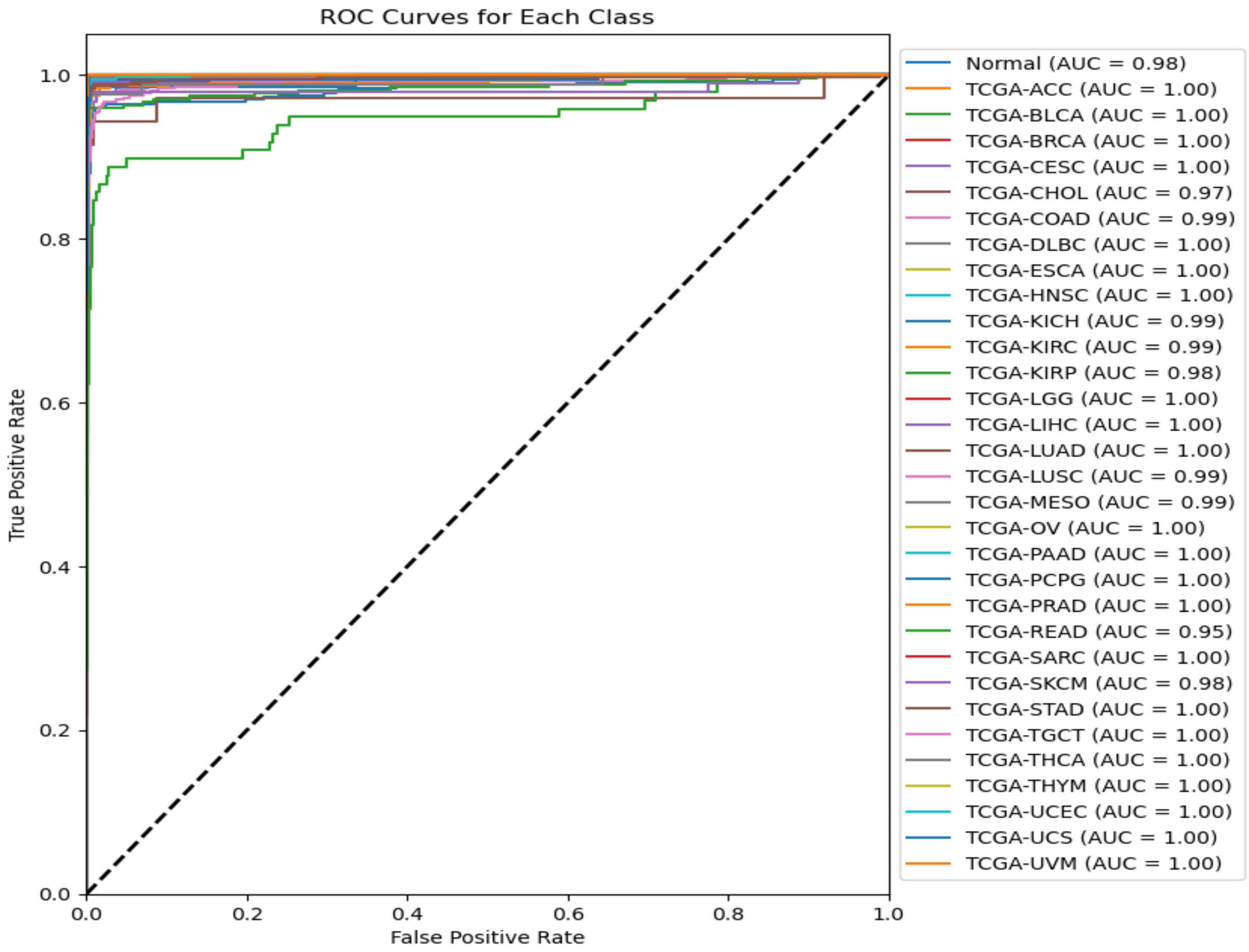

**Figure 4.** ROC Curves and AUC Scores for Multi-Omics Cancer Classification Using GAT Model Across TCGA Cancer Types.

Figure 5 shows the performance comparison of different models. Four key performance metrics are evaluated: accuracy, precision, recall, and F1 score. As depicted in the figure, we observe that the best-performing model is LASSO-MOGAT, which archives the highest accuracy (95.9%), recall (94.45%), and F1 score. Although LASSO-MOGAT outperforms LASSO-MOGCN in terms of accuracy, recall, and F1 score, LASSO-MOGCN gives higher precision which illustrates the model's capability of predicting the positive instances with higher probability.

Table 9 presents a comparison of the performance of different models in cancer classification based on different multi-omics data. The comparison with other models for breast cancer subtypes as well as other cancer classes shows better performance of the proposed LASSO-MOGAT model with an accuracy of 95.90% for 32 cancer classes using mRNA, miRNA, and DNA methylation data. The models by Mostavi et al. (1D-CNN, 2D-Vanilla-CNN, and 2D- Hybrid-CNN) have slightly lower accuracy, where 1D-CNN and 2D-Hybrid-CNN models achieve 95.50% and 95.70%, respectively. The GCNN-PPI model proposed by Ramirez et al. [22] is trained for PPI networks and mRNA data and it is less accurate in comparison with LASSO-MOGAT. Thus, the results show that the inclusion of multiple omics data in the LASSO-MOGAT model enhances the classification performance.

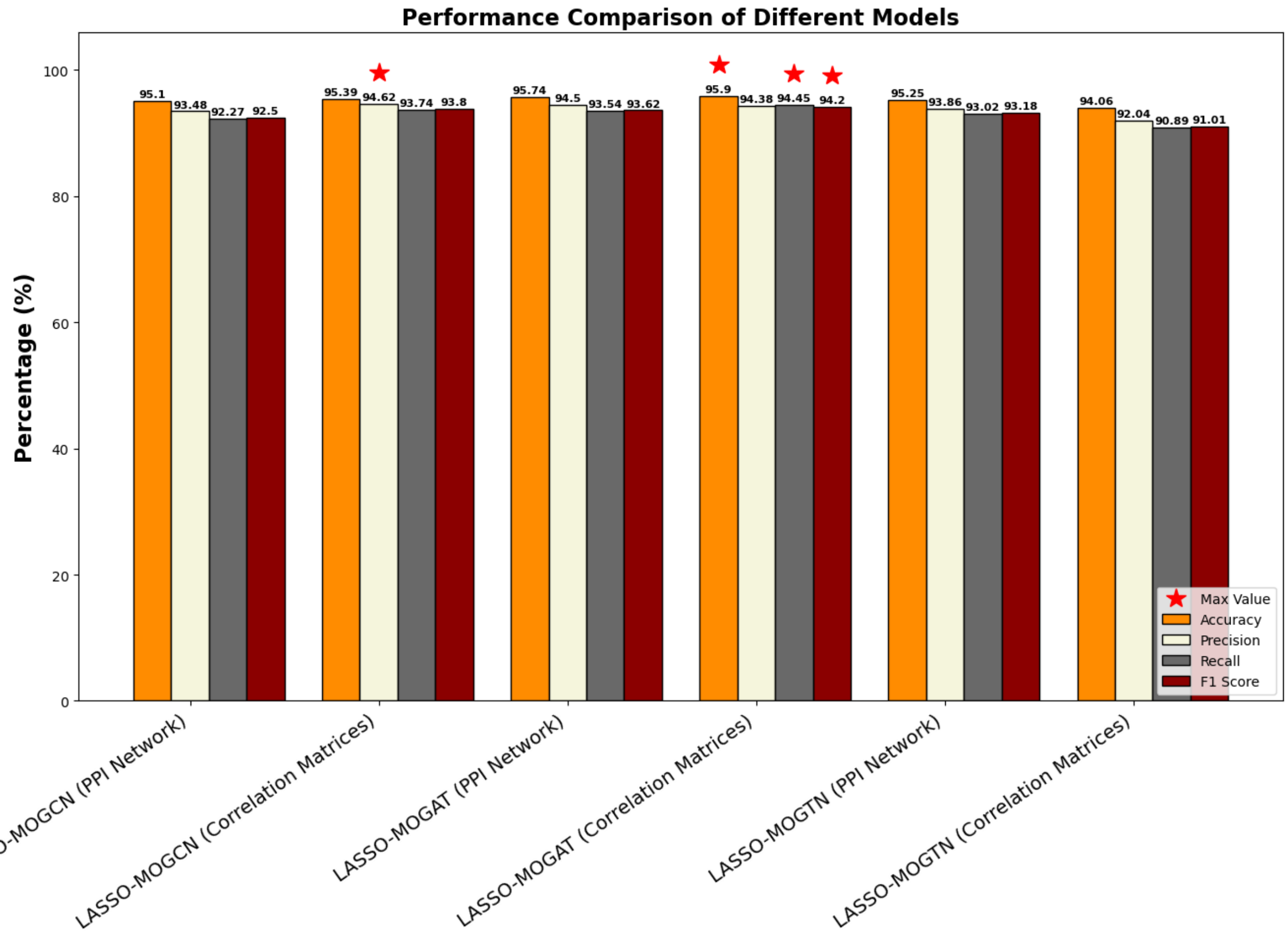

**Figure 5.** Performance comparison of the different models based on multi-omics data types and graph structures.

**Table 9.** Performance metrics for related Single-Omics and Multi-Omics graph methods based on PPI Network.

| **Authors & Models** | **Classes** | **Multi-Omics Data type** | | | **Accuracy** | **Precision** | **Recall** | **F1 Score** |
|---|---|---|---|---|---|---|---|---|
| | | **mRNA** | **miRNA** | **DNA methylation** | Mean ± std | Mean ± std | Mean ± std | Mean ± std |
| **Proposed LASSO-MOGAT** | 32 Classes | √ | √ | √ | **0.9590 ± 0.001** | **0.9438 ± 0.008** | **0.9445 ± 0.009** | **0.9420 ± 0.008** |
| Mostavi et al., 2020 [55] 1D-CNN | 34 Classes | √ | - | - | 0.9550 ± 0.100 | - | - | - |
| 2D-Vanilla-CNN | | √ | - | - | 0.9487 ± 0.040 | - | - | - |
| 2D-Hybrid-CNN | | √ | - | - | 0.9570 ± 0.100 | - | - | - |
| Ramirez et al., 2020 [22] | 34 Classes | √ | - | - | | | | |
| GCNN-PPI graph | | | | | 0.8898 ± 0.883 | 0.8775 | 0.8379 | - |
| GCNN-PPI + singleton graph | | √ | - | - | 0.9461 ± 0.107 | 0.9276 | 0.9219 | - |
| Kaczmarek et al., 2022 [33] **GTN** | 12 Classes | √ | √ | - | 0.9356 ± 0.910 | - | - | - |

## 6. Conclusion

This study presents a novel approach for integrating multi-omics data using graph-based machine learning models, which achieves state-of-the-art performance in classifying 31 cancer types and normal tissues. The method employs LASSO regression to select the most significant features and leverages graph-based architectures such as GCN, GAT, and GTN to model biological interactions within the data. Experimental results demonstrate that the LASSO-MOGAT model based on integrating mRNA, miRNA, and DNA methylation data improves cancer classification outcomes and achieves the highest accuracy of class separation, as well as balanced precision, recall, and F1 score. These results highlight the importance of multi-omics data integration for accurate cancer classification particularly when combined with correlation matrices structures in GAT models. In conclusion, this framework holds promise for improving early cancer diagnosis and advancing personalized medicine, and potentially contributing to more effective patient care.